%% file: complementary_resub_2.tex
\documentclass[aps,prx,groupedaddress,superscriptaddress,notitlepage,twocolumn,nofootinbib]{revtex4-2}
\usepackage{enumerate} 
\usepackage{mathtools}
\usepackage{amsmath,amssymb}

\input{def}

\newcommand{\zdenek}[1]{{\color{red!70!black}#1}}
\newcommand{\lada}[1]{{\color{red!70!black}#1}}
\let\zdenek\relax
\let\lada\relax

\begin{document}

\title{Unifying uncertainties for rotor-like quantum systems}

\author{Ladislav Mi\v{s}ta, Jr.}
\email{mista@optics.upol.cz}
\affiliation{Department of Optics, Palack\'{y} University, 17. listopadu 12, 771 46 Olomouc, Czech Republic}

\author{Matou\v{s} Mi\v{s}ta}
\email{matousmista@seznam.cz}
\affiliation{Gymn\'{a}zium Olomouc-Hej\v{c}\'{\i}n, Tomkova 45, 779 00 Olomouc, Czech Republic}


\author{Zden\v{e}k Hradil}
\email{hradil@optics.upol.cz}
\affiliation{Department of Optics, Palack\'{y} University, 17. listopadu 12, 771 46 Olomouc, Czech Republic}

\begin{abstract}
\zdenek{The quantum rotor represents, after the harmonic oscillator, the next obvious quantum system to study the complementary pair of variables: the angular momentum and the unitary shift operator in angular momentum. Proper quantification of uncertainties and the incompatibility of these two operators are thus essential for applications of rotor-like quantum systems. While angular momentum uncertainty is characterized by variance, several uncertainty measures have been proposed for the shift operator, with dispersion the simplest example. We establish a hierarchy of those measures and corresponding uncertainty relations which are all perfectly or almost perfectly saturated by a tomographically complete set of von Mises states. Building on the interpretation of dispersion as the moment of inertia of the unit ring we then show that the other measures also possess 
the same mechanical interpretation. This unifying perspective allows us to express all measures as a particular instance of a single generic angular uncertainty measure. The importance of these measures is then highlighted by applying the simplest two of them to derive optimal simultaneous measurements of the angular momentum and the shift operator. Finally, we argue that the model of quantum rotor extends beyond its mechanical meaning with promising applications in the fields of singular optics, super-conductive circuits with a Josephson junction  or optimal pulse shaping in the time-frequency domain. Our findings lay the groundwork for quantum-information and metrological applications 
of the quantum rotor and point to its interdisciplinary nature.}
\end{abstract}

\maketitle
\fakepart{Main text}

\section{Introduction}\label{Sec_introduction}

Quantum mechanics imposes rules, which are establishing a sophisticated network of interconnected and subtle conditions, distinguishing the realm of quantum effects from the classical world and provides ultimate bounds on the precision  of involved variables. This is, for instance, the case of the celebrated Heisenberg uncertainty relation for the canonical pair of position and momentum of a (quantum) particle.
The uncertainty relation states that, loosely speaking, the momentum and position of a particle cannot be measured precisely at the same time.
Interestingly, this concept generalizes to other pairs of canonical observables, such as the quadrature operators of the electromagnetic field. 
In this case, the real part of the field plays the role of canonical position, whereas the imaginary part can be considered as the canonical momentum.
These and similar examples have laid the foundations of quantum optics and quantum information processing. 
The important milestones on this route built systematically over one century were: uncertainty relations, coherent (squeezed) states, Einstein-Podolsky-Rosen (EPR) states \cite{Einstein_35}, the Arthurs-Kelly concept of simultaneous detection of non-commuting variables \cite{Arthurs_65,Stenholm_92}, the concept of Bell-like measurement and the representation of quantum state in phase-space pioneered by Wigner \cite{Wigner_32}, Husimi \cite{Husimi_40} or Glauber \cite{Glauber_63}. On the top of those fundamental concepts there are several valuable protocols allowing the processing of quantum information such as quantum teleportation \cite{Braunstein_98} or quantum cryptography \cite{Grosshans_02}. These examples can be cast into the same formal framework since they obey the commutation rule of the Heisenberg-Weyl algebra \zdenek{$[x, p] = i\hbar\openone$},  and observables are related by the continuous Fourier transformation. 

\zdenek{A similarly simple algebraic structure is associated with the quantum rotor, a periodic system with the infinite-dimensional Hilbert state space $L^{2}(-\pi,\pi)$ of functions $f(\phi)$ such that $\int_{-\pi}^{\pi}d\phi|f(\phi)|^2 < \infty$ \cite{Albert_17}. The rotor is fully characterized by the complementary pair of angular momentum $L$ and unitary shift operator in angular momentum $E$ obeying the Lie algebra $\mathfrak{e}$(2) of the Euclidean group in the plane $E(2)$, $[E,L]= E$ \cite{Kastrup_06,Kowalski_02,Raynal_10}, also called the Lerner criterion \cite{Lerner_68}  in connection with quantum phase. In contradistinction with position and momentum, $L$ and $E$ are related by a discrete-continuous Fourier transformation, which can be used to provide solid grounds for quantum  metrology and fully-fledged phase-space representation  \cite{Rigas_08,Kastrup_16,Albert_17,Mista_22}. The developed toolbox finds application in fields as diverse as the theory of damped rotor \cite{Stickler_18} or rotor-based quantum error correction \cite{Vuillot_24}.}  

\zdenek{Recall that historically} the concept of angular momentum and angle  was regarded as unsettled from several reasons, some of which we mention here. \zdenek{Attempts to use the angular position and its variance as an uncertainty measure \cite{Franke-Arnold_04} are  facing to  problems with non-periodicity of the variance, since the result depends on the chosen interval and the variable together with angular momentum do not provide any close algebraic structure. Likewise, the theoretical considerations frequently do not distinguish carefully between angular, phase or time variables what can be also a source of confusions. In simple terms, quantum phase is usually considered as a variable complementary to operator of photon number, whereas time is a parameter conjugated to energy. Phase can be formally linked  either with a Hermitian phase operator  \cite{Barnett_90} or to a generalized measurement \cite{Susskind_64}.
In contrast, the spectrum of the operator $L$ is an unbounded set of integers and the physically meaningful unitary irreducible representations of the group $E(2)$ associated with the $(L,E)$ pair are infinite-dimensional which distinguishes the shift operator from the quantum phase.} As a final remark,  the problem of angle and angular momentum was plagued by possible ambiguities in theoretical formulation; indeed, the quantum mechanics on the circle can be  consistently formulated by means of variables where conjugated variables are combined in rather nontrivial ways  \cite{Kowalski_96}. Here the extremal states are the so called "wrapped Gaussian states" - eigenstates of the  operator $X= E e^{-L-1/2}$, which however does not close on an algebra when combined with its conjugate operator.  \zdenek{A  thorough overview of theoretical concepts linked to angular momentum and shift operator} can be found in \cite{Kastrup_06,Kastrup_16}. For all these reasons the quantum rotor was not considered in applications on an equal footing with the harmonic oscillator,  though  there are  profound analogies in the  mathematical description,  stemming from the similarities between Fourier transformation and  Fourier series \cite{Mista_22}.

\zdenek{One obstacle to the wider application of the quantum rotor is the existence of several different uncertainty measures for the shift operator $E$, none of which is universally accepted as the right one. Furthermore, the uncertainty relations associated with some of the measures do not possess the properties that a good uncertainty relation should have. To illustrate this, let us recap that the uncertainty of the shift operator $E$ was originally quantified by the dispersion $D^2$ \cite{Holevo_11}, the variance of the sine or cosine operators $S$ and $C$, $C-iS=E$ \cite{Carruthers_68}, or the variance of their rotated versions by a fixed angle \cite{Kastrup_06}. Unfortunately, the uncertainty relation for the variance of the angular momentum and the dispersion obtained from the commutation rule \cite{Holevo_11} cannot be saturated \cite{Kowalski_02}. In addition, extremal states saturating analogous uncertainty relation involving the variance of sine (or cosine) operator \cite{Kastrup_06} do not constitute a generalized measurement \cite{Mista_22} and therefore cannot be fully exploited for metrological purposes. However, both shortcomings can be eliminated. Namely, a saturable uncertainty relation for dispersion can be obtained by using the variational approach \cite{Hradil_06}. Similarly, the uncertainty relation with the variance of a rotated sine operator can be made saturable by a tomographically complete set of states if the rotation angle depends suitably on the investigated state. So far, several different state-dependencies of the rotation angle have been proposed in the literature \cite{Hradil_10,Mista_22}, which gave rise to different uncertainties of the shift operator and consequently to different uncertainty relations. Although the relationship between some of these measures is well known, this is not the case for others. The variety of existing measures of uncertainty invites us to clarify the even deeper question of whether there is some unifying view that would link them all, including dispersion. Such a view would not only allow a unified characterization of existing uncertainty measures based on the second moments of the shift operator, but could also be used to design new measures, uncertainty relations, and new types of extremal states.}

\zdenek{In this paper, we develop such a unified framework for the uncertainty measures of the shift operator $E$. First, we formulate a  hierarchy of the known uncertainty measures. Next, inspired by the analogy of dispersion with the moment of inertia of an inhomogeneous ring about an axis perpendicular to its plane and passing through its center of mass \cite{Opatrny_94}  we provide a remarkable unified interpretation of all other measures. This interpretation comes again by considering the measures as moments of inertia of a ring about axes passing through its center of mass, but with rotation axes no longer necessarily limited to the direction perpendicular to the plane of the ring.} The moment of inertia tensor is given by a covariance matrix \zdenek{of the sine and cosine operators}. Although this extension to 3D is apparently ad hoc, it is extremely advantageous from a metrological perspective as it can be used to formulate new tighter bounds if there is an experimental demand for this.

\zdenek{To demonstrate the practicality of the measures studied, we apply them to find the optimal simultaneous measurement of operators $L$ and $E$. Building on the commuting extension of the latter operators \cite{Mista_22} we generalize the uncertainty relations for the dispersion and variance of the rotated sine operator to two systems and minimize them numerically over the product of the corresponding extremal states. What is more, we also find an analytical formula which coincides or very well approximates the numerical bound thereby getting an analogy with the celebrated Arthurs-Kelly relation \cite{Arthurs_65,Stenholm_92} for quantum rotor.}

\zdenek{Finally, we specify several non-mechanical physical models of the quantum rotor complementing its mechanical implementations \cite{Kuhn_17,Stickler_21} and documenting its interdisciplinary role. This includes the superconducting circuits with a Josephson junction, where the rotor complementary pair is given by  number of tunneling Cooper pairs  and the phase over  the Josephson junction  \cite{Koch_07,Vool_17}. Another direction is  shaping of optical pulses in 1D described by the Fourier series, where the complementary pair corresponds to discrete modal index and continuous time variable. As a} last example, beams with orbital angular momentum  \cite{Molina_Terriza_07,Yao_11,Krenn_17} provide  exquisite experimental platform for quantum information processing in higher dimension as demonstrated by several pivotal experiments \cite{Mair_01,Leach_10}. All these examples of quantum rotor-like systems may profit from \zdenek{the theory developed here,} which allows to quantify complementarity of conjugated variables of angular momentum and \zdenek{shift operator}. 

\zdenek{In  Sec.~\ref{Sec_uncertainty_relations} we review concepts for  saturable  uncertainty relations and extremal states for quantum rotor formulated in various contexts over several decades. Special attention is payed to arguments based on a variational principle and a Robertson-like approach.} In Sec.~\ref{Sec_momentum_of_inertia} we unify both approaches  on an equal footing   introducing the  moment of inertia. Importantly such a formulation not only unifies existing approaches but \zdenek{provides} an opportunity to formulate new and tight uncertainties tailored to possible applications. In Sec.~\ref{Sec_simultaneous_measurement} we extend the formulation to the  problem of simultaneous optimal detection of \zdenek{rotor complementary observables, which may find applications beyond} its mechanical interpretation.
Sec.~\ref{Sec_rotor}  provides valuable  examples where the generic theory developed here can be used, including  vortex beams, qubits in super-conducting circuits or optimal pulse shaping. Conclusions in  Sec.~\ref{Sec_conclusion}  summarize all the results  stressing the metrological meaning of extremal states as fully fledged  minimum uncertainty states for  possible metrological applications. Technical calculations related to simplified derivation for generic moment of inertia, ultimate  uncertainty relations for \zdenek{covariance matrix}, and optimal simultaneous measurement are reported in appendices.

\section{Uncertainty relations for quantum rotor}\label{Sec_uncertainty_relations}

The theory \zdenek{developed in this paper} is motivated by the investigation of uncertainties of two complementary variables. The essence of the theory can be \zdenek{qualitatively understood by analogy with position and momentum, for which the minimum} uncertainty states can be derived either by variation  or with the help of 
\lada{the Schr\"{o}dinger-Robertson inequality $\langle(\Delta x )^2\rangle\langle(\Delta p )^2\rangle \ge \frac{\hbar^2}{4}$, where $\langle(\Delta A)^2\rangle=\langle A ^2\rangle-\langle A\rangle^2$ is the variance of an operator $A$.}

\begin{enumerate}[(i)]
\item  Consider  the example of the sum of  weighted variances
\begin{eqnarray} 
 H = \lambda_x  ( x - \langle x \rangle )^2  +   \lambda_p  ( p - \langle p \rangle )^2.
\end{eqnarray}
 We can consider this as a Hamiltonian for a harmonic oscillator. Its  minimum mean value is reached for a state corresponding to the projection into the  lowest eigenvalue state - squeezed vacuum state,   displaced  by all possible values $  \langle x \rangle ,  \langle p \rangle. $  \zdenek{The family of these states, parameterized by $\langle x\rangle$ and $\langle p\rangle$, provides an} over-complete set of \zdenek{(squeezed)} coherent states. 
 
\item  On  the other hand the same solution can be obtained  by  an alternative way using the inequalities
\begin{eqnarray} 
\langle H  \rangle &=&\lambda_x \langle(\Delta x)^2\rangle+\lambda_p\langle(\Delta p)^2\rangle \nonumber\\
&\stackrel{1}{ \geq}&2 \sqrt {\lambda_x \lambda_p}\sqrt{\langle(\Delta x)^2\rangle\langle(\Delta p)^2\rangle} \stackrel{2}{ \geq}  \hbar \sqrt{ \lambda_x \lambda_p}. 
\end{eqnarray}
The inequality $1$ is saturated by matching  the condition $\lambda_x\langle(\Delta x)^2\rangle=\lambda_p \langle(\Delta p)^2\rangle$, whereas inequality $2$ is the well-known Robertson inequality. As before the minimum uncertainty states are   (squeezed) coherent states. 

\end{enumerate}

\lada{Minimum uncertainty states play a crucial role in the concept of simultaneous measurement. In case of quadrature operators the measurement is linked with the commuting pair ${\mathcal X } = x_s + x_a,\, {\mathcal P }= p_s -p_a,\, [ {\mathcal X}, {\mathcal P}]=0$, known also as the EPR  pair, composed of the quadratures of the measured signal ($s$) and the ancilla ($a$).}
 When the measurement is done on a factorized signal and ancillary system, the optimal uncertainty product $\langle(\Delta {\mathcal X})^2\rangle\langle(\Delta {\mathcal P})^2\rangle \ge \hbar ^2  $   reaches its minimum  if \zdenek{variances of local states satisfy $\langle(\Delta x_s)^2\rangle =\langle(\Delta x_a)^2\rangle$ and $\langle(\Delta p_s)^2\rangle= \langle(\Delta p_a)^2\rangle.$ Both constructions (i) and (ii) yield the same results for harmonic oscillator, but differ for the quantum rotor. In the following we summarize the known differences between the harmonic oscillator and the rotor, which are subtle but appear to be essential.}

\zdenek{The quantum rotor is fully characterized by an angular momentum operator and a shift operator, which are in $\phi$-representation given by $L=-i\partial_{\phi}$ and $E=e^{-i\phi}$, and satisfy} the commutation rule \cite{Kastrup_06,Kowalski_02}
\begin{equation}\label{ELcommutator}
[E,L]= E
\end{equation}
of Euclidean algebra $\mathfrak{e}$(2). In order to formulate uncertainties for angular momentum and angular variable  $ ( L, \phi), $  notice that the  standard variance $\langle(\Delta \phi)^2\rangle$ is not  a good uncertainty measure since it is not shift invariant. The statistical dispersion \zdenek{\cite{Holevo_11}}
\begin{equation}
\label{D}
\lada{D^2 = \langle(E^{\dagger}-\langle E^{\dag}\rangle)(E-\langle E\rangle)\rangle  = 1 - |\langle E \rangle |^2}
\end{equation}
represents the simplest choice of the figure of merit including higher order moments of  the angular variable, not just  its variance   $\langle(\Delta \phi)^2\rangle$.  In the following the  angular variable itself will be avoided  in favour of \zdenek{the shift operator.} 

The states minimising the variance of angular momentum under the constraint of fixed dispersion (minimum uncertainty states for  dispersion) can be sought in the form of variational problem \cite{Hradil_06}   for the minimum  eigenvalue of the operator
 \begin{equation}
\left[ L^2 + \mu L  + \frac12(q^* E +  q E^{\dagger})\right]  |\Psi \rangle = a  |\Psi \rangle,
\label{Math}
 \end{equation}
with $\mu, q$  Lagrange multipliers.  The solution is given by  Mathieu functions \cite{McLachlan_47} in $\phi$ - representation - even (cos-like) Mathieu function $\mathrm{ce}_0(\frac{\phi}{2},q)$  for its minimum eigenvalue. The minimum uncertainty state \zdenek{is thus the analogue of the vacuum} state of the harmonic oscillator. Let us denote formally such a state as
$|\mathrm{ce}_0, q\rangle$ for  $\langle L \rangle = 0 $.  Even if  there is no analytical solution in terms of simple algebraic functions, the bound    $B(D)$ can be calculated numerically  \cite{Goldstein_30,Luks_92,Hradil_06} 
 \begin{equation}
 \label{Mathieu}
 \langle(\Delta L)^2\rangle D^2  \geq B(D).
 \end{equation}

Importantly  the Mathieu ground state can  be very closely approximated \lada{\cite{Goldstein_30,Hradil_06} by the von Mises state 
\cite{Bluhm_95,Kastrup_06,Hradil_10}
\begin{eqnarray}\label{vM}
|m,\alpha,\kappa\rangle=\frac{1}{\sqrt{I_0( 2 \kappa) }}\sum_{l\in \mathbb{Z}} e^{i(m-l)\alpha}I_{m-l} (\kappa)|l\rangle
\end{eqnarray}
with $m=0$ and $\alpha=0$. Here $m\in\mathbb{Z}$ is the angular momentum mean, $\alpha$ is an angle,
$\{|l\rangle\}_{l\in\mathbb{Z}}$ are the angular momentum eigenstates, and $I_{n}(z)$ is the modified Bessel function of order $n$ \cite{Watson_44}. 
The parameter  $\kappa\geq0$ represents the spread of angular variable being similar to squeezing for quadrature operators. 
Projection of the state (\ref{vM}) onto the eigenstate of the shift operator $E$, 
$|\phi\rangle=\sum_{l\in\mathbb{Z}}e^{-il\phi}|l\rangle/\sqrt{2\pi}$, 
\begin{eqnarray}
\label{vonM}
  \langle\phi|m,\alpha,\kappa\rangle=\frac{1}{\sqrt{2\pi I_0( 2 \kappa)}}e^{  i m  \phi  + \kappa \cos(\phi- \alpha)  }, 
  \end{eqnarray}
yields the von Mises distribution for the angle
$\phi$: $|\langle\phi|m,\alpha,\kappa\rangle|^2=\exp{[2\kappa \cos(\phi-\alpha)]}/2\pi I_0( 2 \kappa)$, 
which is why the states (\ref{vM}) are referred to as von Mises states.}
Differences between  ground Mathieu and von Mises  states are so small \zdenek{\cite{Hradil_06} 
as to be hardly detectable by current technology and due to this, we can} replace the analytically intractable  bound $B(D)$ by its good approximation in terms of von Mises states. Explicit form can be easily found with the help of   formulae derived in \cite{Mista_22} in parametric form depending on $\kappa$
\begin{eqnarray}
\langle(\Delta L)^2\rangle D^2 = \frac{\kappa}{2} \frac{I_1(2\kappa )}{ I_0(2\kappa )}\left[1  -    \frac{I^2_1(2\kappa)}{ I^2_0(2\kappa)}\right], \\
D^2 = 1  -    \frac{I^2_1(2\kappa)}{ I^2_0(2\kappa)}.
\end{eqnarray}

There are physical reasons why Mathieu  and von Mises  ground states are so closely related. Whereas  the Mathieu ground state minimises the uncertainty product for dispersion and variance of angular momentum, von Mises states  are extremal states for uncertainty product of the angular momentum and \zdenek{the shift operator}.  Indeed 
the commutation relation for rotated sine and cosine operators
\begin{equation}\label{LSalphacommutator}
[S_{\alpha},L{]} = i C_{\alpha},
\end{equation}
$ C_{\alpha} = (e^ {-i \alpha} E^\dag +e^{i \alpha} E)/2,\quad S_{\alpha} = (e^ {-i \alpha} E^\dag - e^{i \alpha} E )/2i$, yields the uncertainty relation
\begin{equation}\label{UR}
\langle(\Delta L)^2\rangle\langle(\Delta S_{\alpha})^2\rangle\ge \frac14 |\langle C_{\alpha}\rangle |^2, 
\end{equation}
which is saturated  by the von Mises states \lada{(\ref{vM})} as
 the solution of the operator equation \zdenek{\cite{Kastrup_06}}
\begin{equation}\label{vMeq}
\left(\Delta L  - i \kappa \Delta S_{\alpha}\right)|\psi \rangle=0.
\end{equation}

The saturable bounds as a function of covariance matrix can be derived by an approach inspired by Ref.~\cite{Hradil_10}. Let us introduce a normalised variable vector 
${\bf x}=(\cos\alpha,\sin\alpha)^{\rm T}$, an unnormalised vector of the first moments ${\bf c}=(\langle C\rangle,\langle S\rangle)^{\rm T}$, \lada{where $C\equiv C_{0}$ and $S\equiv S_{0}$,} and the covariance matrix
\begin{equation}\label{Gamma}
{\bf \Gamma}=\left(\begin{array}{cc}
\langle(\Delta S)^2\rangle & -\frac{1}{2}\langle\{\Delta S,\Delta C\}\rangle \\
-\frac{1}{2}\langle\{\Delta S,\Delta C\}\rangle & \langle(\Delta C)^2\rangle\\
\end{array}\right),
\end{equation}
where $\{A,B\}\equiv AB+BA$ is the anticommutator. Hence, we get
\begin{eqnarray}\label{vectornotation}
\langle C_{\alpha}\rangle={\bf c}^{\rm T}{\bf x},\quad \langle(\Delta S_{\alpha})^2\rangle=\langle[\Delta(S\cos\alpha-C\sin\alpha)]^2\rangle={\bf x}^{\rm T}{\bf \Gamma}{\bf x},
\nonumber\\
\end{eqnarray}
and the uncertainty relation (\ref{UR}) takes the form
\begin{equation}\label{UR1}
\langle(\Delta L)^2\rangle\left({\bf x}^{\rm T}{\bf \Gamma}{\bf x}\right)\ge\frac{1}{4}\left({\bf c}^{\rm T}{\bf x}\right)^2.
\end{equation}
Moving all the quantities dependent on angle $\alpha$, i.e., on the  unit vector ${\bf x}$, to the right-hand side (RHS) of the uncertainty relation, we can maximise the RHS over the vector ${\bf x}$ thereby getting \cite{Hradil_10}, 
\begin{equation}\label{UR2}
\langle(\Delta L)^2\rangle\ge\frac{1}{4}\underset{\|{\bf x}\|=1}{\mbox{max}}\frac{\left({\bf c}^{\rm T}{\bf x}\right)^2}{{\bf x}^{\rm T}{\bf \Gamma}{\bf x}}=\frac{1}{4}\frac{\left({\bf c}^{\rm T}{\bf x}_{\rm o}\right)^2}{{\bf x}_{\rm o}^{\rm T}{\bf \Gamma}{\bf x}_{\rm o}}=\frac14 {\bf c}^{\rm T}{\bf \Gamma}^{-1}{\bf c},
\end{equation}
where ${\bf x}_{\rm o}={\bf J}{\bf \Gamma}{\bf J}^{\rm T}{\bf c}/\sqrt{{\bf c}^{\rm T}{\bf J}{\bf \Gamma}^2{\bf J}^{\rm T}{\bf c}}$, ${\bf J}=i\pmb{\sigma}_{y}$, is the unit vector for which the maximum on the RHS is reached. As a result  the following chain of inequalities can be derived 
\begin{equation}
\label{measures1}
\langle(\Delta L)^2\rangle \stackrel{1}{\geq}\frac{1}{4}\frac{\left({\bf c}^{\rm T}{\bf x}_{\rm o}\right)^2}{{\bf x}_{\rm o}^{\rm T}{\bf \Gamma}{\bf x}_{\rm o}}\stackrel{2}{\geq}\frac14\frac{\|{\bf c}\|^2}{{\bf m}^{\rm T}{\bf \Gamma}{\bf m}}  \stackrel{3} {\geq}  \frac14  \frac{\|{\bf c}\|^2}{\gamma_+}  \stackrel{4}{ \geq}    
 \frac14     \frac{\|{\bf c}\|^2}{D^2},
\end{equation}
where $\|{\bf c}\|^2=|\langle E\rangle |^2$ and ${\bf m} = {\bf c}/\|{\bf c}\|$. Inequality $1$ is given  in (\ref{UR2}), inequality $2$ follows from the Cauchy-Schwarz inequality
\begin{equation}\label{CS}
\left({\bf m}^{\rm T}{\bf \Gamma}{\bf m}\right)\left({\bf m}^{\rm T}{\bf \Gamma}^{-1}{\bf m}\right)\geq1, 
\end{equation}
inequality $3$ trivially estimates the mean value of covariance matrix by the larger eigenvalue $\gamma_+,$   $$\quad \gamma_{\pm}=\frac{1}{2}\left(1-|\langle E\rangle|^{2}\pm|\langle E^2\rangle-\langle E\rangle^2|\right). $$  
\zdenek {The inequality $4$ originally proposed in \cite{Carruthers_68} (see also \cite{Holevo_11}) cannot be saturated except for $D=0$ \cite{Kowalski_02}. This can be seen directly from  the uncertainty relation (\ref{UR}) formulated separately for the variances of the sine and cosine operators. When both uncertainty relations are added,  we get $  \langle(\Delta L)^2\rangle D^2  \ge  \frac14 (1 - D^2).$ Since the uncertainty relations for the sine and cosine operators cannot be saturated simultaneously, the latter inequality is weak and cannot be further exploited in quantum  metrology. Similarly, the  uncertainty relation based on the commutation rule for the angular momentum and the angular position  (like $x $ and $p$) is saturated by truncated Gaussian states \cite{Franke-Arnold_04}. However, this pair of observables does not provide a closed algebra, the uncertainty relation depends on the probability at the boundary, and since the states saturating the relation do not allow to built phase-space representation, they  cannot be exploited for metrology of the quantum rotor. Despite the seeming simplicity of the latter approach, previous results as well as other arguments \cite{Kastrup_06} confirm that the problems associated with the use of the angular position vanish if the shift operator is used to formulate the uncertainty relation for the quantum rotor.}

\zdenek{The denominators ${\bf x}_{\rm o}^{\rm T}{\bf \Gamma}{\bf x}_{\rm o}$, ${\bf m}^{\rm T}{\bf \Gamma}{\bf m}$, $\gamma_{+}$ and $D^2$ appearing on the right-hand sides of the chain of inequalities (\ref{measures1}) represent possible alternative uncertainty measures of the shift operator constructed from elements of the covariance matrix (\ref{Gamma}). Making further use of the inequalities (\ref{measures1}) together with the Cauchy-Schwarz inequality $({\bf c}^{\rm T}{\bf x}_{\rm o})^2\leq\|{\bf c}\|^2$, we can establish the following hierarchy of the uncertainty measures:
\begin{equation}\label{hierarchy}
D^2\geq\gamma_{+}\geq{\bf m}^{\rm T}{\bf \Gamma}{\bf m}\geq{\bf x}_{\rm o}^{\rm T}{\bf \Gamma}{\bf x}_{\rm o}.
\end{equation}
Except for the dispersion} these measures coincide for von Mises states, ${\bf x}_{\rm o}^{\rm T}{\bf \Gamma}{\bf x}_{\rm o}={\bf m}^{\rm T}{\bf \Gamma}{\bf m}=\gamma_{+}$, and all the inequalities $1,2$ and $3$ reduce to equalities. However, the measures are not the same for other than extremal states, since they are forged from different parameters of covariance matrix. The least one is ${\bf x}_{\rm o}^{\rm T}{\bf \Gamma}{\bf x}_{\rm o}$ and it is on the RHS of inequality $1$ but the simplest one is given on the RHS of inequality $2$ and it was heuristically  derived in \cite{Mista_22},  ${\bf m}^{\rm T}{\bf \Gamma}{\bf m}=\langle S_{ -\mbox{arg}\langle E \rangle}^2\rangle$. Note finally that the quantities ${\bf x}_{\rm o}^{\rm T}{\bf \Gamma}{\bf x}_{\rm o}$ and ${\bf m}^{\rm T}{\bf \Gamma}{\bf m}$ are ill-defined for states with $\langle E \rangle=0$, i.e., ${\bf c}=(0,0)^{\rm T}$, in which case we define both of them as $\gamma_{-}=(1-|\langle E^2\rangle|)/2$.
In the next section we show that concept of moment of inertia unifies all the uncertainties analysed above, and  allows to design even tighter uncertainty relations. 
 
 \section{Moment of inertia as the angular uncertainty measure}\label{Sec_momentum_of_inertia}

 Let us review briefly the notion of the moment of inertia tensor from classical mechanics. Consider a rigid body of volume $V$ and mass density $\rho$ that rotates with angular velocity vector $\pmb{\omega}$ along a fixed axis passing through a point $A$. The mass element $dm=\rho dV$ with a position vector ${\bf r}$ relative to the point $A$ has the velocity vector ${\bf  v}  = \pmb{\omega}\times  {\bf r}.$ The angular momentum and the energy of the body then can be cast in the form
\begin{eqnarray}
{\bf L}_{A}&=&\int_{V}[{\bf r}\times{\bf v}]\rho dV={\bf I}_{A}\pmb{\omega}, \\
\lada{W}&=&\frac12\pmb{\omega}^{\rm T}\cdot {\bf L}_{A}= \frac12 {\pmb{\omega}}^{\rm T} {\bf  I}_{A} \pmb{\omega},
\end{eqnarray}
 where ${\bf I}_{A}$ is the moment of inertia tensor with respect to the point $A$, which is 
 defined by the matrix 
\begin{equation}\label{I}
{\bf I }_{A}=\begin{pmatrix}
\int_{V}\left(y^2+z^2\right)\rho dV & -\int_{V}xy\rho dV & -\int_{V}xz\rho dV\\
-\int_{V}xy\rho dV & \int_{V}\left(x^2+z^2\right)\rho dV & -\int_{V}yz\rho dV\\
-\int_{V}xz\rho dV & -\int_{V}yz\rho dV & \int_{V}\left(x^2+y^2\right)\rho dV
\end{pmatrix}.
\end{equation}

\begin{figure} [htb]
    \centering
    \includegraphics[scale=0.65]{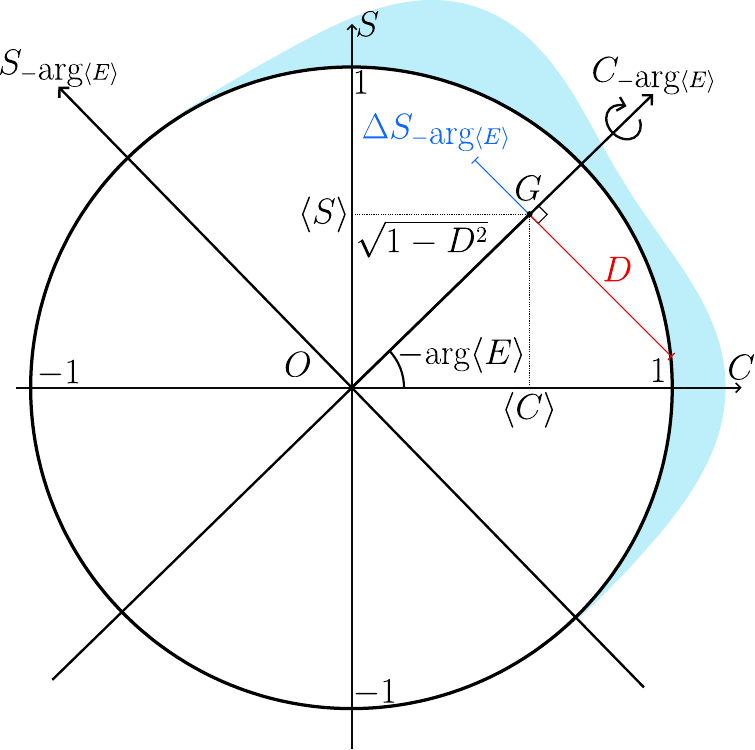}
    \caption{Geometrical and mechanical meaning of angular uncertainty measures for mixture $\rho=0.4\,|0,0\rangle\langle0,0|+0.6\,|0,2\pi/5\rangle\langle0,2\pi/5|$ of two von Mises states with the same parameter $\kappa=5$. The probability density 
    $p(\phi)=\langle\phi|\rho|\phi\rangle$ (light blue area) can be viewed as a linear mass density of an inhomogeneous unit ring with unit mass (black ring). The point $G(\langle C\rangle,\langle S\rangle)$ is the center of mass of the ring. The dispersion $D^2$ (square of the red line segment) is the moment of inertia $M_{{\bf e}_{z}}$, Eq.~(\ref{Mez}), of the ring about an axis perpendicular to its plane and passing through $G$ (see the red axis in Fig.~\ref{ring3D}). The variance $\langle(\Delta S_{-\mbox{arg}\langle E\rangle})^2\rangle$ (square of the blue line segment) is the moment of inertia $M_{\bf e}$, Eq.~(\ref{Me}), of the ring about the  axis $C_{_{-\mbox{arg}\langle E\rangle}}$.}
    \label{ring2D}
\end{figure}

\begin{figure} [htb]
    \centering
    \includegraphics[scale=1.3]{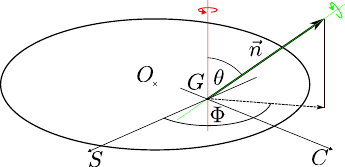}
    \caption{Mechanical interpretation of the dispersion $D^2$, Eq.~(\ref{Mez}), and the generic angular uncertainty measure $M_{\bf n}$, Eq.~(\ref{M}). The dispersion is the moment of inertia of the ring with respect to the red axis. The measure $M_{\bf n}$ is the moment of inertia of the ring about the green axis determined by the unit vector 
    $ {\bf n}=(\sin\theta\cos\Phi, \sin\theta\sin\Phi, \cos\theta )$. See text and caption of Fig.~\ref{ring2D} for details.}
    \label{ring3D}
\end{figure}

Let the body be a unit ring in the $X-Y$ plane, centered at the origin $O$ with the unit mass distributed along the ring with the  angular mass density $p(\phi)$. From the definition (\ref{I}) it  follows that the moment of inertia tensor of the ring about the origin $O$ is given by
\begin{equation}\label{I0}
{\bf I}_{O}=\left(\begin{array}{ccc}
\langle S^2\rangle & -\langle SC\rangle & 0\\
-\langle SC\rangle & \langle C^2\rangle & 0\\
0 & 0 & 1\\
\end{array}\right),
\end{equation}
where we introduced the denotation $\langle S^2\rangle=\langle \sin^2(\phi)\rangle$, $\langle C^2\rangle=\langle \cos^2(\phi)\rangle$, and $\langle SC\rangle=\langle \sin(\phi)\cos(\phi)\rangle$. Recall further the parallel axis theorem connecting the moment of inertia tensor with respect to a point $A$ and the center of mass $G$,   
\begin{equation}\label{Steiner}
{\bf I}_{A}={\bf I}_{G}+m\left(a^2\openone-{\bf a}{\bf a}^{\rm T}\right),
\end{equation}
where ${\bf a}$ is the position vector of the point $G$ relative to the point $A$. For  the ring we have ${\bf a}=(\langle C\rangle,\langle S\rangle,0)^{\rm T}$ and the theorem (\ref{Steiner}) implies
 \begin{eqnarray}
 {\bf I}_{G} = \left(%
\begin{array}{cc}
   {\bf \Gamma} &   0 \\
  0 &  D^2   \\
\end{array}%
\right),
 \end{eqnarray}
 where ${\bf\Gamma}$ is the covariance matrix defined in Eq.~(\ref{Gamma}). Hence, the moment of inertia of the ring about an axis determined by a unit vector ${\bf n}=(n_{x},n_{y},n_{z})^{\rm T}$ and passing through its center of mass $G$ is given by
\begin{equation}\label{Mn}
M_{\bf n}={\bf n}^{\rm T} {\bf  I}_{G} {\bf n}={\bf n}_{xy}^{\rm T}\Gamma{\bf n}_{xy}+D^{2}n_{z}^{2},
\end{equation}
 where ${\bf n}_{xy}=(n_{x},n_{y})^{\rm T}$.
 Due to the formula (\ref{Mn}), the variance $\langle(\Delta S_{\alpha})^2\rangle={\bf x}^{\rm T}{\bf \Gamma}{\bf  x}$, Eq.~(\ref{vectornotation}), can be interpreted   as the  moment of inertia of a ring about an axis passing through its center of mass. The mechanical interpretation and the geometrical meaning of some of the measures is depicted in Figs.~\ref{ring2D} and \ref{ring3D}. Specifically, the dispersion (\ref{D}) is the  moment of inertia of the ring about an axis perpendicular to its plane, i.e., parallel with the unit vector ${\bf e}_{z}=(0,0,1)^{\rm T}$, and passing through its center of mass \cite{Opatrny_94} (see Fig.~\ref{ring2D} and the red axis in Fig.~\ref{ring3D})
\begin{equation}
M_{{\bf e}_{z}} =  1- \langle C_{-\mbox{arg}\langle E\rangle}\rangle^2=D^2.
\label{Mez}
\end{equation}

Analogously, the uncertainty measure  in (\ref{measures1})
 \begin{equation}
M_{\bf e} = \langle S_{-\mbox{arg}\langle E \rangle  }^2\rangle
\label{Me}
\end{equation}
represents the moment of inertia of the ring with respect to the axis connecting the origin $O$ and the center of mass $G$, which is determined by the unit vector ${\bf e}=(\langle C\rangle,\langle S\rangle,0)^{\rm T}/|\langle E\rangle|$ (see axis $C_{-\mbox{arg}\langle E\rangle}$ in Fig.~\ref{ring2D}).

The measures discussed in previous section  are the special cases of moment of inertia with respect to the axis at the center of mass and oriented along the unit vector $ {\bf n}  = ( \sin \theta \cos \Phi, \sin \theta \sin \Phi, \cos \theta )$ (see green axis in Fig.~\ref{ring3D}). A straightforward substitution into the formula (\ref{Mn}) gives the result
\begin{equation}
M_{\bf n} = \langle (\Delta S_{\Phi})^2  \rangle +  \cos^2(\theta)\langle (\Delta C_{\Phi})^2  \rangle = D^2  - \sin^2(\theta)\langle (\Delta C_{\Phi})^2 \rangle. 
\label{M}
\end{equation}
Elementary derivation of generic moment of inertia is given in Appendix A. 
It is also intriguing to note that standard variances used for quantification of uncertainties of position and momentum are nothing but  moments of inertia of a probability distribution along the line (1D) with respect to perpendicular axis in the center of mass.

Note further that the quantity (\ref{M}) is obviously equipped with some of the formal properties of an uncertainty measure as formulated in \cite{Englert_24}. Namely, it is non-negative and for the state-independent angles $\Phi$ and $\theta$ it is obviously well defined and concave. Here, the concavity simply follows from concavity of variance \cite{Hofmann_03} or from the fact that the measure is related to the moment of inertia tensor. To show the latter, consider the convex mixture $\rho=p\rho_{1}+(1-p)\rho_{2}$, and let $I_{G}, I_{G_1}$, and $I_{G_2}$ be the moment of inertia tensors with respect to the center of mass corresponding to the density matrices $\rho, \rho_{1}$ and $\rho_{2}$. Then it is not difficult to show that 
\begin{equation}\label{Icomposition}
{\bf I}_{G}=p{\bf I}_{G_{1}}+(1-p){\bf I}_{G_{2}}\\+p(1-p)\left(R^2\openone-{\bf R}{\bf R}^{\rm T}\right),
\end{equation}
where ${\bf R}=(\langle C\rangle_{\rho_{2}}-\langle C\rangle_{\rho_{1}},\langle S\rangle_{\rho_{2}}-\langle S\rangle_{\rho_{1}},0)^{\rm T}$ is the position vector of the center of mass $G_{2}$ with respect to the center of mass $G_{1}$. Hence, one gets immediately concavity of the measure (\ref{M}),
\begin{equation}
M_{\bf n}(\rho)\geq p M_{\bf n}(\rho_{1})+(1-p)M_{\bf n}(\rho_{2}),
\label{Mconcavity}
\end{equation}
where the Cauchy-Schwarz inequality $R^2\geq|({\bf n}^{\rm T}\cdot{\bf R})|^2$ has been used.

How to understand all those results in the context of possible metrological applications? Moment of inertia is well defined quantity associated with the rotation of a rigid body. 
 According to the inequalities (\ref{Mathieu}) and (\ref{measures1}) moments of inertia, namely \zdenek{$D^2$, $\gamma_{+}$, ${\bf m}^{\rm T}{\bf \Gamma}{\bf m}$ and ${\bf x}_{\rm o}^{\rm T}{\bf \Gamma}{\bf x}_{\rm o}$ appearing in the chain of inequalities (\ref{hierarchy}),} play the role of  the uncertainties  of  the angular variable.   Obviously,  dispersion as  moment of inertia (\ref{Mez})  seems to be the first choice but the Mathieu states are analytically intractable and cannot be fully exploited for further  optimization. On the other hand 
measures constructed from higher order moments (\ref{measures1}) and linked to  moment of inertia (\ref{Me}) may seem to be forged  artificially but give rise to a simple  tractable set of extremal von Mises states, which are effectively indistinguishable from Mathieu states.  In the line with this interpretation the  inequalities  (\ref{Mathieu}) and (\ref{measures1}) cannot be seen as stronger  or weaker since  different measures are involved. However, both   concepts, though slightly different  seem to be equivalent for all practical consequences and, all   measures implied by inequalities (\ref{measures1}) are mimicking dispersion when considering states close to optimal. Of course, differences may appear for non-optimal states due to the differences in  higher order moments. 
  
It follows from this argument that  states  saturating the Robertson's inequality are  ``minimum uncertainty states'' if the concept of uncertainty   is extended to projections of moment of inertia tensor with state-dependent orientation of  axis of rotation:   variance of sine (cosine) operator or dispersion are just extremal  cases of more general formulae (\ref{M}).
Dispersion is in this sense exceptional since the axis is constant (perpendicular to $X-Y$ plane). Why is it worth to have some other uncertainty relations? The answer is simple: They can characterise extremal states under different conditions and may provide tighter uncertainty relations! The tight form of generic uncertainty relations is formulated in Appendix~\ref{Appendix_B}. The extremal states are attributed to solutions of Hill equation \cite{McLachlan_47}, which is a generalization of Mathieu equation.  Here we will just stress  the reasons why this could be of interest: If some  particular  physical platform  of quantum rotor (see examples  below) will allow to identify experimentally  moments of angular variable  $\langle E\rangle$ and $\langle E^2\rangle$,  it could  be valuable to find restrictions implied by quantum mechanics and to formulate stronger uncertainties. We will leave these issues for further research resorting to the simplest opportunity associated with the measures (\ref{Mez}) and (\ref{Me}), which may find direct applications now. 

 \section{Simultaneous detection of complementary variables}\label{Sec_simultaneous_measurement}
 
 The uncertainty relations discussed above have immediate implications for  metrology.  The angular momentum and angular variable can be detected simultaneously in the extended Hilbert space of a signal ($s$) and ancillary ($a$) systems via the commuting pair \cite{Mista_22}
 \begin{eqnarray}
\label{pair} {\cal L } = L_s + L_a,\quad  {\cal E} = E_s E_a^{\dagger}, \quad  [  {\cal L }, {\cal E } ] = 0.
\end{eqnarray}
 It is plausible to assume that commuting pair (\ref{pair}) represents the optimal scheme for  any purpose.  Here we address just the strategy  based on minimizing the uncertainties. For metrology on  signal state the ancillary system is controlled independently from the signal system and the global state is factorised $|\psi \rangle_{sa} = |\varphi\rangle_{s}|\chi\rangle_{a}$. Setting the ancillary system to some fiducial state $|f\rangle$, which will be specified later, leads to an over-complete POVM in signal space projecting onto the states $D(m, \phi)|f\rangle$ and satisfying the completeness condition
\begin{equation}
\sum_{m\in\mathbb{Z}}\int_{-\pi}^{\pi}\frac{d\phi}{2\pi}  D(m, \phi)  |f \rangle \langle f | D^{\dagger} (m,\phi) = \openone,
\label{closure}
\end{equation}
where $D(m, \phi) = e^{-iL \phi} E^{-m}$  is the  displacement operator with an arbitrary phase factor being omitted.  The choice of the fiducial state dictated by the optimality can be either the ground Mathieu  state $|ce_0,q\rangle$,  von Mises state $|0,0,\kappa \rangle$, or other optimal state discussed in previous Section.  
The figures of merit used to quantify uncertainties of the angular momentum and the angular variable for the commuting pair (\ref{pair}) can be cast in the form
 \begin{eqnarray}
\langle(\Delta {\cal L } )^2\rangle&=&\langle(\Delta L_s)^2\rangle  +\langle(\Delta L_a)^2\rangle,\label{calL}\\
\mathcal{D}^2&=&1-|\langle\mathcal{E}\rangle|^{2}=| \langle E_a\rangle |^2 D_{s}^{2}+D_{a}^{2},\label{calD}\\
\langle(\Delta\mathcal{S})^2\rangle&=&|\langle E_{a}^2\rangle|\langle(\Delta S_s)^2\rangle+\langle(\Delta S_a)^2\rangle.\label{calS}
\end{eqnarray}
\zdenek{The operator on the left-hand side of Eq.~(\ref{calS}) is defined as 
$\mathcal{S}=(e^{-i\beta}\mathcal{E}^\dag - e^{i\beta}\mathcal{E})/2i$, where 
$\beta=\mbox{arg}\langle E_{a}\rangle-\mbox{arg}\langle E_{s}\rangle$, which implies 
$\langle\mathcal{S}\rangle=0$. For the second moment we have 
$\langle\mathcal{S}^2\rangle=[1-|\langle E_{s}^2\rangle ||\langle E_{a}^2\rangle |\cos(\delta_{s}-\delta_{a})]/2$, with $\delta_{j}=\arg\langle E_{j}^{2}\rangle-2\arg \langle E_{j}\rangle$, $j=s,a$. To suppress the unwanted influence of the ancilla on the measurement we assume it to be in such a state that $\arg \langle E_a \rangle  = 0$ and $\arg \langle E_a^{2} \rangle  = 0$ \cite{Mista_22}. This gives $\delta_{a}=0$, the moment $\langle\mathcal{S}^2\rangle$ of the composite system exhibits the same 
angular dependence on the moments of the signal state as the signal moment   
$\langle S_{s,-\mbox{arg}\langle E \rangle  }^2\rangle=(1-|\langle E_{s}^2\rangle |\cos\delta_{s})/2$, and $\langle(\Delta\mathcal{S})^2\rangle$ reduces to the RHS of Eq.~(\ref{calS}).}

Simultaneous detection  exhibits added noise in both  angular momentum and angular variable, but the latter one is penalized  by an extra multiplicative factor $|\langle E_a\rangle |^{2}$ or $|\langle E_a^2\rangle |$, Eqs.~(\ref{calD}) and (\ref{calS}), respectively, a  consequence of the fact  that the angular variable is always measured with respect to a  reference. 

Full analysis of optimal simultaneous detection is a delicate task, which depends on  constraints.  Just to get the flavour  we will specify   two opposite scenarios, namely to optimize  ancilla state for a given signal state or conversely, to optimize signal state for a given ancilla state. The first task is more involved since it requires to consider measurements which depend on the measured signal - a situation which frequently happens  when Quantum Fisher information is considered. Here we address the second task, which has a straightforward metrological meaning answering the question what signal is optimally detected by a given apparatus (ancilla).  Detailed discussions will be done here for 
 dispersion,  the analogous arguments for  $\langle(\Delta\mathcal{S})^2\rangle$ can be found in Appendix~\ref{Appendix_C}. The analysis is facilitated by introducing the vector notation
\begin{eqnarray}\label{ld}
    \pmb{\cal l}&=&( \Delta L_s, \Delta L_a)^{\rm T},\quad {\bf d}_{ij} = (|\langle E_{i}\rangle|  D_{j}, D_{i})^{\rm T},
\end{eqnarray}
where $i,j=a,s$, and $\Delta A\equiv\sqrt{\langle(\Delta A)^2\rangle}$. 
The rationale behind the definition of 
$ {\bf d}_{ij} $  stems from the fact that separate uncertainties $ \langle(\Delta {\cal L } )^2\rangle$ and $\mathcal{D}^2$ can be interpreted as norms of vectors, for which the  Cauchy-Schwarz  inequalities will  represent saturable bounds with several possible branches. 
As a result the uncertainty product can be assessed as
\begin{align}\label{PicalD}
&\Pi_{\mathcal D}=\langle(\Delta {\cal L } )^2\rangle \mathcal{D}^2=||\pmb{\cal l}||^2 ||{\bf d}_{ij}||^2\stackrel{1}{\geq}|(\pmb{\cal l}^{\rm T}\cdot {\bf d}_{ij})|^2\nonumber\\
&=
\begin{cases}
\,\,\,(|\langle E_s\rangle|\Delta L_{s}D_{a}+\Delta L_{a}D_{s})^2  & \text{for\,\, $ij=sa$};
   \\
\,\,\,(|\langle E_a\rangle|\Delta L_{s}D_{s}+\Delta L_{a}D_{a})^2
   \\
\,\,\,\stackrel{2}{\geq}\left(|\langle E_a\rangle |\sqrt{B_s}+\sqrt{B_a}\right)^2
  &\text{for\,\, $ij=as$},
   \end{cases}
\end{align}
where \lada{we used the abbreviation $B_s \equiv B(D_s), B_a \equiv B(D_a)$. Here,} inequality $1$ follows from the Cauchy-Schwarz inequality, whereas inequality $2$ is a consequence of the uncertainty relation (\ref{Mathieu}). The analysis hinges partially on numerical analysis due to the dependence on the state-dependent factor $\langle E_a\rangle.$ 
\lada{The results together with the bound $B(D)$ in inequality~(\ref{Mathieu}) and its approximation for von Mises states are summarized in Fig.~\ref{figD}.}
In Appendix~\ref{Appendix_C1} we show  that the numerically found optimal value of the uncertainty product $\Pi_{\mathcal D}$ can be well approximated for ancilla prepared in the Mathieu state as  ${\cal B}_{\mathcal D}$, 
\begin{align}
\label{calBD}
&{\cal B}_{\mathcal D}=\left\{\begin{array}{lll}
\left(|\langle E_a\rangle |\sqrt{B_s}+\sqrt{B_a}\right)^2 & \!\textrm{for}\! & \hspace{-0.1cm}D_{a,{\rm int}}^2 >D_{a}^2\geq0;\\
\langle(\Delta L_{a})^2\rangle & \!\textrm{for}\! & \hspace{-0.1cm} 1\geq D_{a}^2\geq D_{a,{\rm int}}^2,
\end{array}\right.
\end{align}
 \begin{figure} [htb]
    \centering
    \includegraphics[scale=0.45]{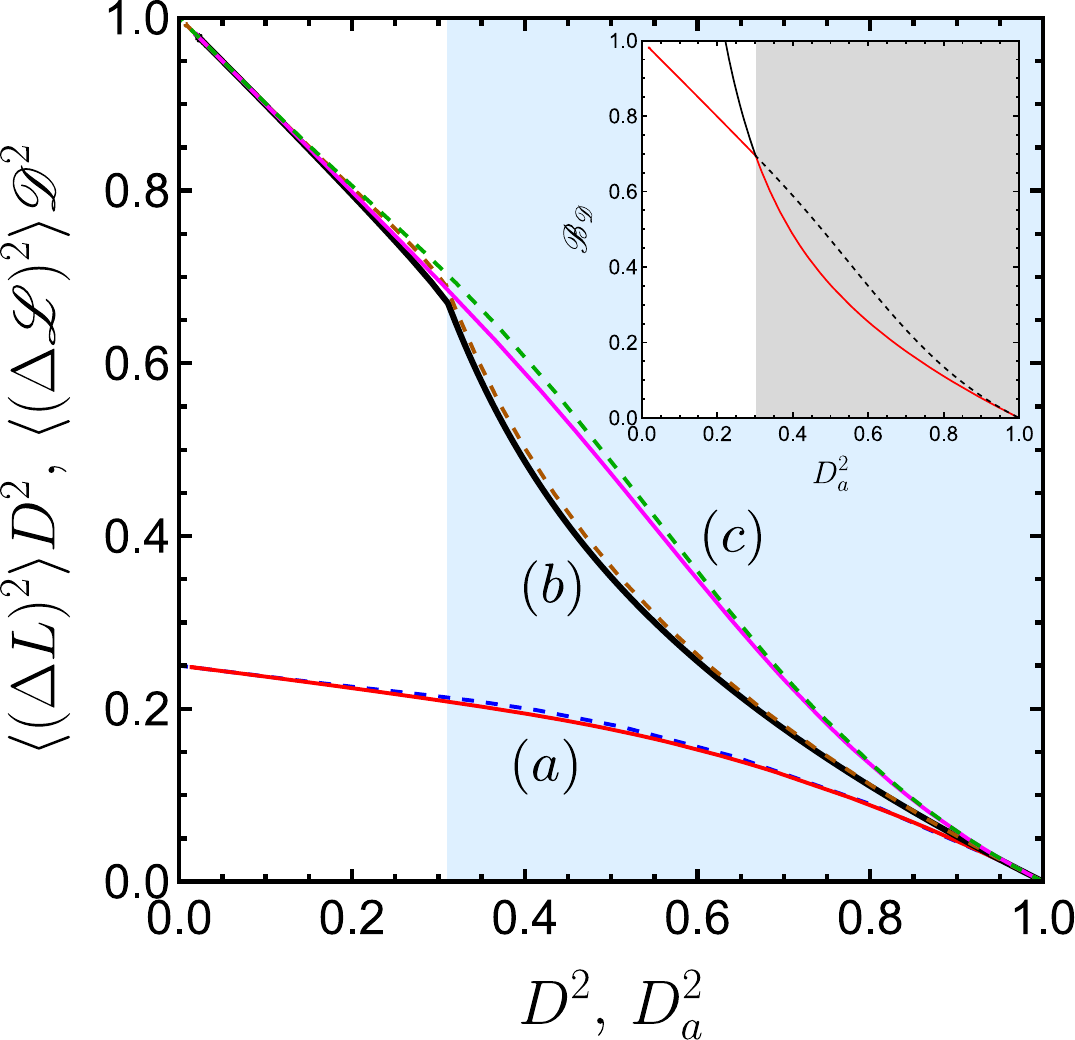}
    \caption{Uncertainty product $\langle(\Delta L)^2\rangle D^2$  as a function of   dispersion $D^2$ $(a)$  and $\Pi_{\mathcal{D}}=\langle(\Delta{\cal L})^2\rangle{\cal D}^2$ as a function of  $D_a^2$  $(b)$, $(c)$. The products are plotted for the extremal 
    Mathieu states $|\mathrm{ce}_0, q\rangle$ (solid lines) and the von Mises states $|0,0,\kappa\rangle$ (dashed lines). The pair of lines $(a)$ shows that the von Mises states  approximate the optimal Mathieu states very closely and this correspondence propagates into simultaneous measurement under various strategies. The pair of lines $(b)$ corresponds to the numerical solutions with optimally matched Mathieu and von Mises states, and in case of $(c)$ the signal is matched to ancilla according to the conditions saturating the Cauchy-Schwarz inequality. The inset displays the function ${\cal B}_{\cal G}$ characterizing minimal $\Pi_{\mathcal{D}}$ (solid red line) and the remaining parts of its two branches (black lines). The branches give analytical arguments explaining the numerical solution.}
    \label{figD}
\end{figure}
 \begin{figure} [htb]
    \centering
    \includegraphics[scale=0.45]{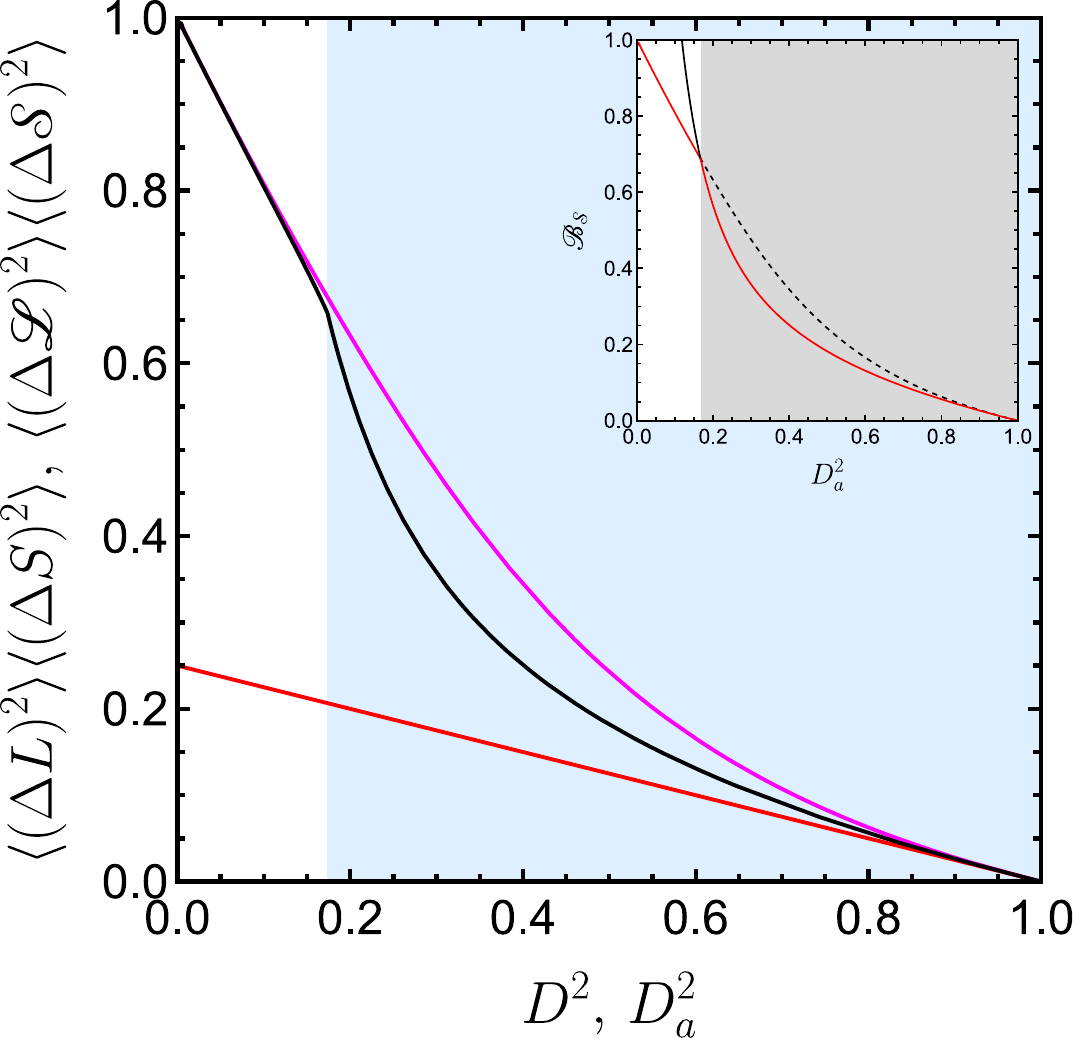}
    \caption{Uncertainty product $\langle(\Delta L)^2\rangle\langle(\Delta S_{-\mbox{arg}\langle E\rangle})^2\rangle$ as a function of dispersion $D^2$ (red line) and $\Pi_{\mathcal{S}}=\langle(\Delta{\cal L})^2\rangle\langle(\Delta\mathcal{S})^2\rangle$ as a function of  $D_a^2$. The products are plotted for the extremal von Mises states $|0,0,\kappa\rangle$.  The black line corresponds to the numerical solutions with optimally matched von Mises states, and the magenta line depicts the case when the signal is matched to ancilla according to the conditions saturating the Cauchy-Schwarz inequality. The inset displays the function ${\cal B}_{\cal S}$ characterizing minimal $\Pi_{\mathcal{S}}$ (solid red line) and the remaining parts of its two branches (black lines).}
    \label{figS}
\end{figure}
where $D_{a,{\rm int}}^{2}\doteq 0.3$. The derivation was done here for dispersion and Mathieu  extremal states but  analogous formulation done  for the uncertainty measure $ \langle(\Delta\mathcal{S})^2\rangle$, Eq.~(\ref{calS}), and  von Mises states (see Fig.~\ref{figS}) shows the consistency of both the measures. In Appendix~\ref{Appendix_C2} we further show  that the optimal numerical bound is very well approximated by the function 
\begin{align}
\label{calBS}
&{\cal B}_{\cal S}=\left\{\begin{array}{lll}
\frac{1}{4}\left(\sqrt{|\langle E_a^{2}\rangle|}|\langle E_s\rangle|+|\langle E_a\rangle|\right)^2 & \!\textrm{for}\! & \hspace{-0.1cm}\tilde{D}_{a,{\rm int}}^2>D_{a}^2\geq0;\\
\frac{\langle(\Delta L_{a})^2\rangle}{2} & \!\textrm{for}\! & \hspace{-0.1cm} 1\geq D_{a}^2\geq \tilde{D}_{a,{\rm int}}^2,
\end{array}\right.
\end{align}
where $\tilde{D}_{a,{\rm int}}^2\doteq 0.167$ and the parameters of the signal and ancilla are matched by the condition $\kappa_{s}=\sqrt{I_2(2\kappa_{a})/I_0(2\kappa_{a})}\kappa_{a}$ \cite{Mista_22}.
This result has clear interpretation as condition matching  the squeezing  of signal and ancillary systems. Notice that for $x$ and $p$ quadrature operators of the harmonic oscillator both signal and ancillary systems are  squeezed equally.  

 Disregarding the  subtle differences between von Mises and Mathieu extremal states and  optimal conditions  for  matching  signal and ancillary systems, the "almost optimal" simultaneous detection of complementary variables can be  seen as detection of  the von Mises signal state  projected onto sequence of von Mises detector states (\ref{closure}) - a result fully  analogous to the heterodyne detection for harmonic oscillator when squeezed state can be measured by projecting into squeezed (Gaussian) states. However the structure of optimal conditions is much richer for quantum rotor with promises for future experimental realisations.

\section{Quantum rotor in applications}\label{Sec_rotor}
  
There are several distinct and important    physical platforms   for implementing quantum rotor in the current technology.     Detailed analysis of specific examples is beyond the scope of this  contribution. Here we just point out  common features of all the examples, namely that  uncertainties for complementary variables represent the first step towards optimal metrology.

As the first example let us mention  system with cylindrical symmetry-vortex beams.   Uncertainties for dispersion and variance of angular momentum were analyzed theoretically and   verified experimentally in  Ref. \cite{Hradil_06}.  It is intriguing to note that  \zdenek{the shift operator} can be constructed on two modes  of transversal  field associated with annihilation operators $a_1, a_2$   as so called "feasible" phase  \cite{Hradil_92}
\begin{eqnarray}
L= a^{\dagger}_1  a_1 - a_2^{\dagger }a_2,   \quad    E  =  \sqrt{     \frac{a_1 + a_2^{\dagger}   }{ a_1^{\dagger}  +  a_2  }    }. 
\end{eqnarray}
As shown by Ban \cite{Ban_91}, there is an equivalent \zdenek{representation of the shift operator,
\begin{eqnarray}
E = \sum_{m=0}^{\infty}\sum_{n=-\infty}^{\infty }| n-1,m\rangle\rangle\langle\langle m,n|, \end{eqnarray}
where
\begin{eqnarray}
  | n, m \rangle \rangle = \theta(n) |m+n , m \rangle  + \theta(-n-1) |m , m-n \rangle
\end{eqnarray}
is the so called relative number state. Here $\theta(n) = 1 , n \ge 0,$ and $\theta(n) = 0 , n \le  0$ and both representations are unitarily equivalent \cite{Ban_94}.}
These concepts come from formal considerations of quantum phase operator and  provide exact  (and intrinsically nonlinear) link  between algebra of the harmonic oscillator and the quantum rotor. 
  
  Less obvious is  the link between the problem of   optimal shaping of the  pulse in the time-frequency domain  within the model of   quantum rotor.  For the sake of simplicity we assume  t-dependent signal  on the $ 2\pi$ window in dimensionless coordinates,  where $t$ plays the role  of  angular variable.  Outside this interval let us   fulfill   the periodicity condition $ \psi (t + 2\pi)   = \psi(t). $     Such signals can be    represented  by  discrete Fourier series  in the  selected time window
 \begin{equation}
 \psi(t) = \sum_{n= -\infty}^{\infty} a_n  e^{i t n}.
 \label{pulse}
 \end{equation}
 If we define operators in $t$-representation  as  $  L = -i \partial_t , \quad  E = e^{-i t} , $ we easily identify the commutation  relation for $\mathfrak{e}$(2) algebra  (\ref{ELcommutator}) with all the metrological consequences.   In contrast with the standard formulation where $x, p$  are related by Fourier transformation, the  discrete    $n$   and continuous $t$-variables are parameters in Fourier series.  This problem will be addressed in a subsequent publication. 
 
  Contemporary quantum computing platforms  for performing quantum operations are based  on  circuits with Josephson junctions \cite{Arute_19,Wu_21}.
  Superconductors behave like  macroscopic quantum mechanical systems. Only two quantities are required to describe the physics of a Josephson junction: the number imbalance of electrons $n$ (Cooper pairs)  and  $\theta $  is the relative phase between the two superconductors. 
   The  circuits   built with  superconducting components  can carry currents without resistance because the carriers of the charge - electrons or holes near the Fermi energy level - are  creating the Cooper pairs   behaving like  macroscopic coherent states as explained by BCS theory.  Such a state can be described by complex-valued  order parameter, the phase of which is essential  for the  physics  of super-conducting qubits \cite{Makhlin_01,Rasmussen_21}.  
     The standard explanation in   solid state physics relies on the canonical pair   
\zdenek{$
 "[ \hat \theta,  \hat n ]" = i\openone , 
$}
though  it is known that this form is mathematically not rigorous due to the  periodicity of angular variable \cite{Devoret_04,Vool_17,Kwon_21}. 
The quotation marks are indicating the potential problems: the number of Cooper pairs $n$ tunneling through the barrier in Josephson junction can be negative  while  $\theta$   linked with gauge phase over the barrier is periodic,  bearing the striking similarity with phase-number  operators.  However, the relation $  e^{i\theta} n  e^{-i\theta} = n-1$ is free of those problems but this is nothing else but the algebraic expression for Euclidean algebra  $\mathfrak{e}$(2), Eq.~(\ref{ELcommutator}),  for the \zdenek{shift operator} $  E= e^ { - i\theta}.   $  

There are still  other analogies with  uncertainty relations for the quantum rotor.
 Under the conditions of super-conductivity   only  Cooper pairs tunnel coherently in the superconducting junction, and the system is described by the Hamiltonian of island-base qubit with capacitance coupled to Josephson junction  \cite{Makhlin_01}  called also Cooper pair box 
 \begin{equation}
 H= 4 E_C (n- n_g)^2 -E_J  \cos \theta, 
 \label{hamiltonian}
 \end{equation}
 where $E_C$ is the charging energy and  $ E_J$ is the Josephson junction energy. As pointed out in \cite{Devoret_04}, the Cooper box is to quantum circuit physics what the hydrogen atom is to atomic physics.
 But this Hamiltonian  of a quantum rotor  \cite{Koch_07} 
 is the same as the extremal equation (\ref{Math}) for states minimising uncertainty for the variance of angular momentum and dispersion (\ref{Mathieu}). In other words, projections into the (displaced) ground state of the quantum rotor have the same meaning for quantum tomography of super-conducting qubits as projections into  (displaced) ground state of harmonic oscillator, i.e., projections into squeezed states of electromagnetic field. Since Mathieu states can be very closely   approximated by von Mises states the theory presented here provides deep analogy between  quantum metrology of harmonic oscillator and quantum rotor.   This is facilitated by the generalised measurement of commuting pair (\ref{pair}), where the  signal and ancillary loops should be coupled to more complex circuits allowing to detect super-imposed  (or subtracted)  complementary variables from both  loops in similar way as it is done in quantum optics with homodyne or heterodyne detection.
Practical and ambitious  goal of theoretical analysis presented in this research  could be  the design of  optimised detection  and diagnostic schemes  for circuits  built on the basis of a Josephson junction, which provides platforms  for   charge-, phase- or fluxons  super-conducting qubits  \cite{Blais_21,Rasmussen_21}.

\section{Conclusion}\label{Sec_conclusion}
 
\zdenek{In this paper we developed the theory of uncertainty measures and corresponding uncertainty relations for the angular momentum and the unitary shift operator of the quantum rotor. We have shown that all relevant measures can be hierarchically ordered and that they can all be interpreted as moments of inertia of an inhomogeneous unit ring with respect to different axes of rotation passing through its center of mass. This unifying perspective allowed us to find one general uncertainty measure of the shift operator, from which all other measures follow as special cases. The general measure can provide even tighter uncertainty relation compared to its particular instances, and the corresponding extremal states can be obtained as solutions of the Hill equation. The extreme states associated with the particular uncertainty measures are given by or are very well approximated by von Mises states. These states provide a tractable representation suitable for metrology of the quantum rotor at ultimate quantum limits, which closely resembles the squeezed-state representation for the harmonic oscillator. To demonstrate practical utility of the developed formalism we applied it to find the optimal simultaneous measurements for the angular momentum and the shift operator. First, we generalized uncertainty relations involving dispersion and variance of the rotated sine operator to two-rotor commuting extension of this complementary pair. Next, we minimized the generalized uncertainty relation over the product of the corresponding minimum uncertainty states thereby getting the analogy of the Arthurs-Kelly relation for the quantum rotor. Last but not least, we identified two non-mechanical physical systems suitable for realizations of the Euclidean algebra of the quantum rotor encompassing pulses in the time-frequency domain and super-conducting circuits with a Josephson junction. The presented results have the potential to accelerate development of metrological and quantum information applications of rotor-like quantum systems.}

\begin{acknowledgments}

The authors are indebted to Jaroslav \v{R}eh\'{a}\v{c}ek and Hubert de Guise for valuable comments. We acknowledge financial support from H2020-FETOPEN-2018-2019-2020-01 StormyTune.

\end{acknowledgments}

\bibliography{bibrotor}
\appendix

\section{Elementary derivation of moment of inertia for arbitrary axis}\label{Appendix_A}

Here we give elementary  and straightforward  derivation of the moment of inertia (\ref{M}) without using tensor calculations. 
 Assume the coordinate system of the unit ring, where $X$ axis goes through its center of mass,  vectors of  axis of rotation ${\bf n} = (\cos \Phi \sin \theta, \sin \Phi \sin \theta, \cos \theta) $ and position on the unit ring  at the $X-Y$ plane  $ {\bf e}  = (\cos \varphi, \sin \varphi, 0) .$  The moment of inertia of the ring with respect to the axis at origin  is  given as 
 \begin{equation}
     M_0 = \langle \sin^2(\vartheta) \rangle_{\varphi},
 \end{equation}
where $ \cos \vartheta  = {\bf n}{\bf e}  = \cos(\Phi-\varphi)  \sin \theta.$  The moment of inertia with respect to parallel axis in the center of mass  is given by the parallel axis theorem as 
\begin{equation}
 M =    M_0  - |\langle C \rangle |^2 \cos^2(\theta).
 \end{equation}
Using  the relation $D^2 = 1 - |\langle C \rangle |^2  $ and $ D^2 =  \langle (\Delta S_{\Phi})^2 \rangle +  \langle (\Delta C_{\Phi})^2 \rangle $
we get finally the expression where the moment of inertia is given as weighted sum of both cosine and sine variances
\begin{eqnarray}
 M =   D^2   - \langle (\Delta C_{\Phi})^2 \rangle \sin^2(\theta)  \\
 =  \langle (\Delta S_{\Phi})^2 \rangle  +  \langle (\Delta C_{\Phi})^2 \rangle \cos^2 (\theta).
 \end{eqnarray}
 Notice that \zdenek{there are two extremal measures - dispersion is the largest whereas  $\gamma_{-}$ is the smallest.}

\section{The tight uncertainty relations  for generic moment of inertia}\label{Appendix_B}

There are three parameters of the covariance matrix (\ref{Gamma}), e.g., its trace, determinant and relative phase  between moments   or just moments $\langle E \rangle,  \langle E^2 \rangle $  (apart of an overall arbitrary phase) which  can be exploited for quantification of  the "noise" associated with the covariance matrix.  

The first formulation is motivated by a similar approach as in inequality (\ref{Mathieu}) seeking  the  minimum of the variance of angular momentum under the constraint of fixed values of moments $ \langle E \rangle $  and $ \langle E^2 \rangle $ 
\begin{equation}
   \langle(\Delta L)^2\rangle \ge B (\langle E \rangle,\langle E^2 \rangle).
    \label{B1}
\end{equation}
The extremal states and the uncertainty can  be found as  solution of the ground state of the Hamiltonian 
 \begin{equation}
\left( L^2 + \mu L  + \frac12q^* E + \frac12 q E^{\dagger}   +  \frac12 r^* E^2 +  \frac12r E^{ 2\dagger}    \right)  |\Psi \rangle = a  |\Psi \rangle,
 \end{equation}
$\mu, q, r$ being Lagrange multipliers.  Since we are seeking the solution for $\langle L \rangle = 0 $ and zero phase, we  might set $\mu = 0$ and in $\phi$-representation this tends to Hill equation \cite{McLachlan_47}, similarly to previously discussed problem of Mathieu function, 
\begin{equation}
- \psi''(\phi)  +   [q    \cos(\phi) + r \cos(2\phi-\beta) ] \psi(\phi) = a \psi(\phi),
 \end{equation}
$\beta = \arg[ \langle E^2\rangle\langle E \rangle^{*2} ] .$ Inequality (\ref{B1})
does not have the form of uncertainty relations since it does not contain  angular uncertainty term.

Uncertainty relations based on the moment of inertia can be considered as minimum of the  product
$\langle(\Delta L)^2\rangle M_{\bf n}  $  for the fixed values of  $\theta, \Phi$ under the constraint of  fixed  $D^2$ and $\langle (\Delta C_{\Phi})^2 \rangle.$ Extremal states are again solutions of Hill equation. Finally, the product should be minimised with respect to  $\theta, \Phi $ and these parameters should be identified as  state dependent quantities analogously to construction adopted in derivation of inequalities (\ref{measures1}) saturated by von Mises states.

Notice that von Mises state - a solution of Eq.~(\ref{vMeq}), can be also cast as a special case of (factorized) Hill equation  corresponding to the ground state of the Hamiltonian 
\begin{equation}\label{vMH}
\left(\Delta L  + i \kappa \Delta S_{\alpha}\right)\left(\Delta L  - i \kappa \Delta S_{\alpha}\right)|\psi \rangle=0.
\end{equation}

\section{Saturable bounds for simultaneous measurement}\label{Appendix_C}

In this section we analyse the inequalities for the uncertainty products  
 \begin{eqnarray}
 \Pi_{\mathcal D}&=&\langle(\Delta {\cal L } )^2\rangle \mathcal{D}^2,\label{PiD}\\
\Pi_{\mathcal S}&=&\langle(\Delta {\cal L } )^2\rangle\langle(\Delta {\cal S} )^2\rangle,\label{PiS}
\end{eqnarray}
where the uncertainties on the RHS are defined in Eqs.~(\ref{calL})-(\ref{calS}) 
of the main text. First, we analyse the product of uncertainties with dispersion and then move on to the case containing the variance of the sine operator.

\subsection{Bounds for uncertainty product $\Pi_{\mathcal D}$}\label{Appendix_C1}

At the outset it is convenient to introduce the vectors 
\begin{eqnarray}\label{ld}
\pmb{\cal l}_{ij}&=&( \Delta L_i, \Delta L_j)^{\rm T},\quad {\bf d}_{ij} = (|\langle E_{i}\rangle|  D_{j}, D_{i})^{\rm T},
\end{eqnarray}
where $i,j=s,a$, and $\Delta A\equiv\sqrt{\langle(\Delta A)^2\rangle}$. This allows 
us to write 
 \begin{eqnarray}
\langle(\Delta {\cal L } )^2\rangle&=&||\pmb{\cal l}_{sa}||^2=||\pmb{\cal l}_{as}||^2,\label{l}\\\
\mathcal{D}^2&=&||{\bf d}_{sa}||^2=||{\bf d}_{as}||^2,\label{d}
\end{eqnarray}
and express the product (\ref{PiD}) in the following four different ways:
\begin{equation}\label{PiDcases}
\Pi_{\mathcal D}=||\pmb{\cal l}_{ij}||^2 ||{\bf d}_{kl}||^2,
\end{equation}
where $(ij,kl)=(sa,sa),(sa,as),(as,sa),(as,as)$. Each combination leads to a different bound whereby the bounds corresponding to the combinations $(as,sa)$ and $(as,as)$ are not better than the other two bounds. If we restrict ourselves to cases corresponding to combinations $(sa,sa)$ and $(sa,as)$ and introduce for simplicity the denotation $\pmb{\cal l}\equiv\pmb{\cal l}_{sa}$ we get the following estimates
\begin{align}\label{PD}
&\Pi_{\mathcal D}=||\pmb{\cal l}||^2 ||{\bf d}_{ij}||^2\stackrel{1}{\geq}|(\pmb{\cal l}^{\rm T}\cdot {\bf d}_{ij})|^2\nonumber\\
&=
\begin{cases}
\,\,\,(|\langle E_s\rangle|\Delta L_{s}D_{a}+\Delta L_{a}D_{s})^2\equiv{\cal A}_{1}  & \text{for\,\, $ij=sa$};
   \\
\,\,\,(|\langle E_a\rangle|\Delta L_{s}D_{s}+\Delta L_{a}D_{a})^2
   \\
\,\,\,\stackrel{2}{\geq}\left(|\langle E_a\rangle |\sqrt{B_s}+\sqrt{B_a}\right)^2\equiv{\cal A}_{2}
  &\text{for\,\, $ij=as$}.
   \end{cases}
\end{align}

The inequality $1$ is a consequence of the Cauchy-Schwarz inequality and it is saturated for $\pmb{\cal l}=\lambda{\bf d}_{ij}$, i.e., for $ij=sa$ when 
\begin{eqnarray}\label{condsa}
\Delta L_s D_{s} = |\langle E_{s} \rangle| \Delta L_a D_{a}
\end{eqnarray}  
holds, whereas for $ij=as$ when 
\begin{eqnarray}\label{condas}
\Delta L_s D_{a} = |\langle E_{a} \rangle| \Delta L_{a} D_{s}
\end{eqnarray}  
is satisfied. The inequality $2$ then follows from uncertainty relations (\ref{Mathieu}) of the main text for the signal and ancilla, where we used the abbreviation $B_s \equiv B(D_s), B_a \equiv B(D_a)$. This justifies the metrological role of extremal states since saturation is achieved for signal and ancilla prepared in Mathieu state with the uncertainties matching the condition (\ref{condas}). 

If we limit ourselves to cases where the ancilla is prepared in the Mathieu state $|ce_0,q_{a}\rangle_{a}$, we have $\Delta L_a D_{a}\leq 1/2$ (see red line in Fig.~\ref{figD} of the main text), and the signal state then would have to satisfy $\Delta L_s D_{s}\leq |\langle E_{s}\rangle|/2$. However, this would  contradict to unsaturable inequality $\Delta L_s D_{s}\geq |\langle E_{s}\rangle|/2$ following from inequality $4$ of (\ref{measures1}). Thus the condition (\ref{condsa}) can only be satisfied for $|\langle E_{s}\rangle|=0$ implying $D_{s}^2=1$ and $\langle(\Delta L_s)^2\rangle=0$, and the optimal signal state is an angular momentum eigenstate $|l\rangle_{s}$. The branch ${\cal A}_{1}$ in Eq.~(\ref{PD}) then reduces to ${\cal A}_{1}=\langle(\Delta L_a)^2\rangle$ and it lies below the branch ${\cal A}_{2}$ of the same equation. The two branches intersect if $(\Delta L_{a}-\sqrt{B_{a}})/|\langle E_{a}\rangle|=\sqrt{B_{s}}$, which happens for $q_{a,{\rm int}}\doteq 9.29$ giving $D_{a,{\rm int}}^{2}\doteq 0.3$. For larger $D_{a}$ the branch ${\cal A}_{2}$ is, on the other hand, less than the branch ${\cal A}_{1}$. Thus assuming the ancilla prepared in the Mathieu state we find the uncertainty product $\Pi_{\mathcal{D}}$, Eq.~(\ref{PD}), to be bounded from below by
\begin{align}
\label{calB}
&{\cal B}_{\mathcal D}=\left\{\begin{array}{lll}
\left(|\langle E_a\rangle |\sqrt{B_s}+\sqrt{B_a}\right)^2 & \!\textrm{for}\! & \hspace{-0.1cm}D_{a,{\rm int}}^2 >D_{a}^2\geq0;\\
\langle(\Delta L_{a})^2\rangle & \!\textrm{for}\! & \hspace{-0.1cm} 1\geq D_{a}^2\geq D_{a,{\rm int}}^2.
\end{array}\right.
\end{align}

It is of interest to compare the obtained bounds with the minimum of the uncertainty product (\ref{PiD}) for the Mathieu states $|\mathrm{ce}_0, q_{s}\rangle_{s}|\mathrm{ce}_0, q_{a}\rangle_{a}$. Numerical minimization of the product over $q_{s}$ at fixed $q_{a}$ yields on a restricted interval of dispersions $D_{a}^2$ a little lower bound compared to the bound (\ref{calB}). The resulting function is depicted by a solid black line in Fig.~\ref{figD} of the main text. Similar to the bound (\ref{calB}) the obtained curve again contains a sharp point, but now for a slightly different $q_{a,{\rm sh}}\doteq8.7$, for which $D_{a,{\rm sh}}^2\doteq 0.31$. In the region $1\geq D_{a}^2\geq D_{a,{\rm sh}}^2$ the obtained numerical curve coincides exactly with the second branch $\langle(\Delta L_{a})^2\rangle$ of the bound (\ref{calB}), whereas for $D_{a,{\rm sh}}^2 >D_{a}^2\geq0$ the numerically found minimal uncertainty product lies a little below the first branch $(|\langle E_a\rangle |\sqrt{B_s}+\sqrt{B_a})^2$ (c.f. solid black line and solid magenta line in white area of Fig.~\ref{figD}). For comparison, in Fig.~\ref{figD} we also plotted by the dashed orange line the minimal uncertainty product for the product $|0,0,\kappa_{s}\rangle_{s}|0,0,\kappa_{a}\rangle_{a}$ of the von Mises states and by the dashed green line the uncertainty product for von Mises states satisfying the condition (\ref{condas}). We see that in both cases the obtain curves are again only a little worse than the black and magenta line for the Mathieu states. The observed subtle differences between separate solutions as well as Mathieu and von Mises states will become important when the experimental techniques will be able to distinguish among them. 

\subsection{Bounds for uncertainty product $\Pi_{\mathcal S}$}\label{Appendix_C2}

Moving to the uncertainty product (\ref{PiS}), assume for simplicity the signal 
and ancilla to be prepared in the von Mises states $|n,\alpha,\kappa_{s}\rangle_{s}$ and $|0,0,\kappa_{a}\rangle_{a}$, respectively. Let us further introduce the vector
\begin{eqnarray}\label{s}
{\bf s}_{ij}&=&\left(\sqrt{|\langle E_{i}^2\rangle|} \Delta S_j, \Delta S_i\right)^{\rm T}, 
\end{eqnarray}
where
\begin{eqnarray}\label{DeltaSj}
\Delta S_j&=&\sqrt{\langle(\Delta S_j)^2\rangle}=\sqrt{\langle S_{j,-\mbox{arg}\langle E_{j}\rangle}^2\rangle}\nonumber\\
&=&\sqrt{\frac{1}{2}\left(1-|\langle E_{j}^2\rangle|\right)}.
\end{eqnarray}
Similar to Eqs.~(\ref{d}) and (\ref{PiDcases}) we can then write 
 \begin{eqnarray}
\langle(\Delta\mathcal{S})^2\rangle&=&||{\bf s}_{sa}||^2=||{\bf s}_{as}||^2,\label{DcalS}
\end{eqnarray}
and 
\begin{equation}\label{PiScases}
\Pi_{\mathcal S}=||\pmb{\cal l}_{ij}||^2 ||{\bf s}_{kl}||^2,
\end{equation}
where $(ij,kl)=(sa,sa),(sa,as),(as,sa),(as,as)$. Again, there are four ways of how we can decompose the uncertainty product (\ref{PiS}) but only combinations $(sa,sa)$ and $(sa,as)$ are relevant. Consequently, we get for the uncertainty product (\ref{PiS}) the following inequalities 
\begin{align}\label{PS}
&\Pi_{\mathcal S}=||\pmb{\cal l}||^2 ||{\bf s}_{ij}||^2\stackrel{1}{\geq}|(\pmb{\cal l}^{\rm T}\cdot {\bf s}_{ij})|^2\nonumber\\
&=
\begin{cases}
\,\,\,\left(\sqrt{|\langle E_s^2\rangle|}\Delta L_{s}\Delta S_{a}+\Delta L_{a}\Delta S_{s}\right)^2\equiv{\cal C}_{1} & \text{for $ij=sa$};
   \\
\,\,\,\left(\sqrt{|\langle E_a^2\rangle|}\Delta L_{s}\Delta S_{s}+\Delta L_{a}\Delta S_{a}\right)^2
   \\
\,\,\,\stackrel{2}{\geq}\frac{1}{4}\left(\sqrt{|\langle E_a^{2}\rangle|}|\langle E_s\rangle|+|\langle E_a\rangle|\right)^2\equiv{\cal C}_{2}
  &\text{for $ij=as$}.
   \end{cases}
\end{align}
Here, to get inequality $1$ we used the Cauchy-Schwarz inequality and the equality is obtained if 
and only if $\pmb{\cal l}=\lambda{\bf s}_{ij}$, i.e., for $ij=sa$ when 
\begin{eqnarray}\label{condSsa}
\Delta L_s \Delta S_{s} = \sqrt{|\langle E_{s}^2\rangle|}\Delta L_a \Delta S_{a}
\end{eqnarray}  
is obeyed, whereas for $ij=as$ when 
\begin{eqnarray}\label{condSas}
\Delta L_s \Delta S_{a}=\sqrt{|\langle E_{a}^2\rangle|}\Delta L_a \Delta S_{s}
\end{eqnarray}  
is fulfilled. The inequality $2$ then comes from the uncertainty relation \cite{Mista_22}
\begin{eqnarray}\label{URappendix}
\langle(\Delta L_{j})^2\rangle\langle(\Delta S_j)^2\rangle\geq\frac{1}{4}|\langle E_j\rangle|^2,
\end{eqnarray}
where $j=s,a$. Substituting here from \cite{Mista_22}
\begin{equation}
\langle(\Delta L_{j})^2\rangle=\frac{\kappa_{j}}{2}\frac{I_1(2\kappa_{j})}{I_0(2\kappa_{j})},\quad
\langle(\Delta S_j)^2\rangle=\frac{1}{2\kappa_{j}}\frac{I_1(2\kappa_{j})}{I_0(2\kappa_{j})},
\end{equation}
we find that in terms of the parameters $\kappa_{s}$ and $\kappa_{a}$ the condition (\ref{condSsa}) reads as 
\begin{equation}\label{condSBessel1}
\frac{I_1(2\kappa_{s})}{I_0(2\kappa_{s})}=\sqrt{\frac{I_2(2\kappa_{s})}{I_0(2\kappa_{s})}}\frac{I_1(2\kappa_{a})}{I_0(2\kappa_{a})}.
\end{equation}  
Likewise, the condition (\ref{condSas}) boils down to \cite{Mista_22}
\begin{equation}\label{condSBessel2}
\kappa_{s}=\sqrt{\frac{I_2(2\kappa_{a})}{I_0(2\kappa_{a})}}\kappa_{a}.
\end{equation}  
Numerical analysis reveals that for a given $\kappa_{a}$ the condition 
(\ref{condSBessel1}) is satisfied by $\kappa_{s}=0$. This gives 
$\langle(\Delta L_{s})^2\rangle=0$ and it is again optimal to measure the angular momentum eigenstates $|l\rangle_{s}$. What is more, $\langle(\Delta S_s)^2\rangle=1/2$ and the branch ${\cal C}_{1}$ reduces to ${\cal C}_{1}=\langle(\Delta L_{a})^2\rangle/2$. The latter branch intersects with the second branch ${\cal C}_{2}$, Eq.~(\ref{PS}), for $\kappa_{a}$ satisfying the following equation   
\begin{equation}\label{Sintersection}
\frac{\sqrt{2}\Delta L_{a}-|\langle E_{a}\rangle|}{\sqrt{|\langle E_{a}^2\rangle|}}=|\langle E_{s}\rangle|,
\end{equation}  
where the parameter $\kappa_{s}$ is given by the RHS of the condition (\ref{condSBessel2}).
Upon solving previous equation we find that the branches ${\cal C}_{1}$ and ${\cal C}_{2}$ intersect for $\kappa_{a,{\rm int}}\doteq 3.018$ which gives $\tilde{D}_{a,{\rm int}}^2\doteq 0.167$.
Since for $D_{a}^2<\tilde{D}_{a,{\rm int}}^2$ the branch ${\cal C}_{2}$ lies below the branch ${\cal C}_{1}$, whereas for $D_{a}^2>\tilde{D}_{a,{\rm int}}^2$ it is the other way around, we find the uncertainty product (\ref{PiS}) to be greater or equal to  
\begin{align}
\label{calBSappendix}
&{\cal B}_{\cal S}=\left\{\begin{array}{lll}
\frac{1}{4}\left(\sqrt{|\langle E_a^{2}\rangle|}|\langle E_s\rangle|+|\langle E_a\rangle|\right)^2 & \!\textrm{for}\! & \hspace{-0.1cm}\tilde{D}_{a,{\rm int}}^2>D_{a}^2\geq0;\\
\frac{\langle(\Delta L_{a})^2\rangle}{2} & \!\textrm{for}\! & \hspace{-0.1cm} 1\geq D_{a}^2\geq \tilde{D}_{a,{\rm int}}^2.
\end{array}\right.
\end{align}
The function (\ref{calBSappendix}) is depicted by the red line in the inset of 
Fig.~\ref{figS} of the main text. Its first branch is displayed by the magenta line 
in the main figure, as well as in the inset, where it consists of the dashed black line in the gray area and the solid red line in the white area. The second branch of the bound (\ref{calBSappendix}) is depicted only in the inset and it consists of the solid red line in the gray area and solid black line in the white area. The value of the dispersion where the function (\ref{calBSappendix}) exhibits the sharp point is $\tilde{D}_{a,{\rm int}}^2\doteq 0.167$ and it is depicted by the vertical border between the white and gray area.  

By the numerical minimization of the uncertainty product (\ref{PiS}) over the parameter 
$\kappa_{s}$ at fixed $\kappa_{a}$ we get the black line in Fig.~\ref{figS} of the main text.
The line again possesses a sharp point but now for a slightly different value of $\kappa_{a,{\rm sh}}\doteq2.897$, for which $\tilde{D}_{a,{\rm sh}}^2\doteq 0.174$. Comparison with the numerically minimized uncertainty product (\ref{PiS}) reveals that in the interval $1\geq D_{a}^2\geq \tilde{D}_{a,{\rm sh}}^2$ the second branch $\langle(\Delta L_{a})^2\rangle/2$ is equal to the minimal uncertainty product, whereas for $\tilde{D}_{a,{\rm sh}}^2>D_{a}^2\geq0$ the first branch lies slightly above the minimal uncertainty product (c.f. black line and magenta line in the white area of Fig.~\ref{figS} of the main text). 

\end{document}

%% file: def.tex
\usepackage[latin1]{inputenc}
\usepackage{amsthm}
\usepackage{amssymb}
\usepackage{amsmath}
\usepackage{bbold}
\usepackage{bbm}
\usepackage{nonfloat}
\usepackage{braket}
\usepackage{dsfont}
\usepackage{mathdots}
\usepackage{mathtools}
\usepackage{enumerate}
\usepackage{csquotes}
\usepackage{stmaryrd}
\usepackage[cal=boondox]{mathalfa}
\usepackage{graphicx}
\usepackage{stackengine}
\usepackage{scalerel}
\usepackage{array}
\usepackage{makecell}
\newcolumntype{x}[1]{>{\centering\arraybackslash}p{#1}}
\usepackage{tikz}
\usetikzlibrary{shapes.geometric, arrows, arrows.meta, decorations.pathreplacing, decorations.markings, decorations.pathmorphing, angles, quotes}
\usepackage{booktabs}
\usepackage{xfrac}
\usepackage{siunitx}
\usepackage{centernot}
\usepackage{comment}
\usepackage{marvosym}
\usepackage{chngcntr}
\usepackage{times,txfonts}

\tikzstyle{box} = [rectangle, rounded corners, minimum width=1.618cm, minimum height=1cm,text centered, draw=black]
\tikzstyle{arrow} = [thick,->,>=stealth]

\newtheorem{thm}{Theorem}
\newtheorem*{thm*}{Theorem}
\makeatletter
\newcommand{\setthmtag}[1]{
  \let\oldthethm\thethm
  \renewcommand{\thethm}{#1}
  \g@addto@macro\endthm{
    \addtocounter{thm}{-1}
    \global\let\thethm\oldthethm}
  }
\makeatother

\newtheorem*{prop*}{Proposition}

\newtheorem*{lemma*}{Lemma}

\newtheorem*{cor*}{Corollary}

\newtheorem*{cj*}{Conjecture}

\newtheorem*{Def*}{Definition}

\makeatletter
\def\thmhead@plain#1#2#3{%
  \thmname{#1}\thmnumber{\@ifnotempty{#1}{ }\@upn{#2}}%
  \thmnote{ {\the\thm@notefont#3}}}
\let\thmhead\thmhead@plain
\makeatother

\theoremstyle{definition}

\newcommand{\bb}{\begin{equation}}
\newcommand{\bbb}{\begin{equation*}}
\newcommand{\ee}{\end{equation}}
\newcommand{\eee}{\end{equation*}}

\DeclareMathAlphabet{\pazocal}{OMS}{zplm}{m}{n}

\newcommand{\lsmatrix}{\left(\begin{smallmatrix}}
\newcommand{\rsmatrix}{\end{smallmatrix}\right)}

\stackMath

\stackMath

\makeatletter
\newcommand*\rel@kern[1]{\kern#1\dimexpr\macc@kerna}
\newcommand*\widebar[1]{%
  \begingroup
  \def\mathaccent##1##2{%
    \rel@kern{0.8}%
    \overline{\rel@kern{-0.8}\macc@nucleus\rel@kern{0.2}}%
    \rel@kern{-0.2}%
  }%
  \macc@depth\@ne
  \let\math@bgroup\@empty \let\math@egroup\macc@set@skewchar
  \mathsurround\z@ \frozen@everymath{\mathgroup\macc@group\relax}%
  \macc@set@skewchar\relax
  \let\mathaccentV\macc@nested@a
  \macc@nested@a\relax111{#1}%
  \endgroup
}

\DeclareSymbolFontAlphabet{\mathcalorig}{symbols}

\newcommand{\fakepart}[1]{
 \par\refstepcounter{part}
  \sectionmark{#1}
}

\counterwithin*{equation}{part}
\counterwithin*{thm}{part}
\counterwithin*{figure}{part}